\documentclass[12pt,preprint]{aastex}
\usepackage{natbib}
\usepackage{multirow}
\usepackage{upgreek}
\usepackage[colorlinks=true, pdfstartview=FitV, linkcolor=blue,citecolor=blue, urlcolor=blue]{hyperref}
\usepackage{wasysym}
\usepackage{txfonts}
\usepackage{gensymb}
\usepackage{morefloats}
\usepackage{threeparttable}
\slugcomment{version \today}
\shorttitle{Two Compton-Thick Active Nuclei in Arp 220?}
\shortauthors{Paggi et al.}
\begin{document}
\title{Two Compton-Thick Active Nuclei in Arp 220?}
\author{Alessandro Paggi\altaffilmark{1}, Giuseppina Fabbiano\altaffilmark{1}, Guido Risaliti\altaffilmark{1,2}, Junfeng Wang\altaffilmark{1} and Martin Elvis\altaffilmark{1}}
\affil{\altaffilmark{1}Harvard-Smithsonian Center for Astrophysics, 60 Garden St, Cambridge, MA 02138, USA: \href{mailto:apaggi@cfa.harvard.edu}{apaggi@cfa.harvard.edu}\\
\altaffilmark{2}INAF-Arcetri Observatory, Largo E, Fermi 5, I-50125 Firenze, Italy}
\begin{abstract}
Narrow-band spectral imaging with sub-pixel resolution of the
\textit{Chandra}-ACIS archival observation of the ULIRG merger Arp 220 
strongly suggests two Compton thick nuclei, spatially coincident with the 
infrared and radio emitting nuclear clusters, and separated by 1'' (\(\sim 365\) 
pc at a distance of 76 Mpc). These previously undetected highly obscured AGNs 
- West (W) and East (E) - are imaged, and separated from neighboring sources, 
in the 6-7 keV band, where the Fe-K lines dominate the emission. The western 
nucleus is also detected at energies above 7 keV. We estimate Fe-K equivalent 
width \(\sim 1\) keV or possibly greater for both sources, and observed 2-10 
keV luminosities \(L_X < 3.2\times{10}^{40}\mbox{ erg}\mbox{ s}^{-1}\) (W) 
and \(<1.3\times {10}^{40}\mbox{ erg}\mbox{ s}^{-1}\) (E). From the observed 
Fe-K lines luminosities{, and assuming on the basis of the \textit{XMM-Newton} spectrum that 40\% of this may be from the 6.4 keV component,} we evaluate 2-10 keV intrinsic luminosities 
{\(L_X \sim 1\times {10}^{42}\mbox{ erg}\mbox{ s}^{-1}\) (W) and 
\(L_X \sim 0.4\times {10}^{42}\mbox{ erg}\mbox{ s}^{-1}\) (E)}. The inferred 
X-ray luminosity is at least a factor of 3 higher than that expected from a pure 
starburst with the bolometric luminosity of Arp 220. For a typical AGN SED the 
bolometric luminosities are {\(5.2\times {10}^{43}\mbox{ erg}\mbox{ 
s}^{-1}\) (W) and \(2\times {10}^{42}\mbox{ erg}\mbox{ s}^{-1}\) (E).} 
\end{abstract}
\keywords{galaxies: active --- galaxies: individual (Arp 220) --- galaxies: Seyfert --- galaxies: interactions --- X-rays: galaxies}

\section{INTRODUCTION}\label{intro}
The M-\(\sigma\) relation \citep[e.g.,][]{1998AJ....115.2285M} has suggested
that the evolution of galaxies and super-massive nuclear black holes (SMBHs)
are linked. Both the stellar population and the SMBH of a galaxy are thought to
grow and evolve by merging of smaller gas-rich galaxies and their nuclear
SMBHs \citep{2005Natur.433..604D,2008ApJ...682L..13H}. During this process,
the SMBH may be ``buried" by thick molecular gas, which feeds the SMBH at
high rates, causing the birth of an obscured Compton Thick 
\citep[CT,][]{1999ApJ...522..157R,2006ApJ...648..111L} Active Galactic Nucleus
(AGN). CT AGNs are characterized in the X rays by a hard high energy
continuum, a ``reflection" flat continuum in the \(\sim 2-10\) keV range, and
a very high Equivalent Width (EW) \(\gtrsim 1\) keV 6.4 Fe-K\(\alpha\) line 
\citep[e.g.,][]{1997A&A...325L..13M,2000MNRAS.318..173M}.
Examples of this merger-driven evolution are given by the pairs of nuclei 
discovered in the 6.4 keV Fe-K line with \textit{Chandra} in the merger infrared 
(IR) luminous galaxy NGC 6240 \citep{2003ApJ...582L..15K} and in the CT AGN
NGC 3393 \citep{2011Natur.477..431F}.

At a distance of 76 Mpc \citep{1998ApJS..119...41K}, Arp 220 (IC 4553/4) is 
both a merger, and the nearest Ultra-luminous IR Galaxy 
\citep[ULIRG;][]{1987ApJ...320..238S,1996ARA&A..34..749S}. Near IR 
high-resolution (0.1'') NICMOS-HST imaging identifies the two nuclear regions 
of the merging galaxies, which are coincident with the two components of a 
double radio source \citep{1995ApJ...454..745B,1998ApJ...492L.107S}. At a 
separation of 0.98" (361 pc at a distance of 76 Mpc
\footnote{In the following, we adopt the standard flat cosmology with 
\(\Omega_\Lambda = 0.73\) and \(H_0 = 70 \mbox{ km}\mbox{ 
s}^{-1}\mbox{ Mpc}^{-1}\) \citep{2011ApJS..192...18K}.}), these nuclei are 
closer together than the nuclei of NGC 6240 (\(\sim 690\) pc separation, 
\citealt{2003ApJ...582L..15K}), and therefore should be subject to 
even stronger gravitational interaction and possible accretion. Indeed, the 
presence of an AGN with a contribution to the bolometric luminosity between 
\(\sim 5\) and \(\sim 25\%\), and a best estimate of 18\%, is suggested by the 
Spitzer Mid-IR spectrum of the central 8" region of Arp 220 
(\citealt{2009ApJS..182..628V,2010MNRAS.405.2505N}; more recent 
\textit{Herschel} results \citep{2011ApJ...743...94R} and modeling of the 
nuclear spectra \citep{2012MNRAS.tmp..247C} agree with this conclusion.
The presence of a maser \citep{2009A&A...493..481A} and a rotating massive 
molecular disk \citep{2007A&A...468L..57D} suggests a massive nuclear black 
hole in the west nucleus of Arp 220.

A high resolution study with a \(\sim 57\) ks Chandra ACIS observation
\citep[obsid 869;][]{2002ApJ...581..974C} failed to secure the firm 
identification of nuclear AGN emission, reporting the presence of three hard 
X-ray sources in the region, of which two (X-1 and X-4, see their Figure 2) are 
near, but not coincident with, the nuclear radio sources. The X-ray spectrum 
extracted from the central \(\sim 2\)'' region showed complexity, with a 
possible hard component and Fe-K line. Subsequent \textit{XMM-Newton} 
observation detected Fe-K line emission centered at 6.7 keV with EW 
\(\sim 1.9\) keV \citep{2005MNRAS.357..565I} suggesting highly 
photoionized, low-density gas illuminated by a hidden CT AGN. {A re-
analysis of the \textit{Chandra} data \citep{2011ApJ...729...52L} only manages 
to set an upper limit on the Fe-K\(\alpha\) EW assuming a 6.4 keV line energy}. 
The nature of the X-ray emission of Arp 220 is therefore still elusive.

In this paper we re-examine the question of the X-ray AGN emission of Arp 
220, by means of sub-pixel imaging of Chandra ACIS data in narrow spectral 
ranges. This technique has been used successfully to study crowded emission 
regions of nearby Seyferts (e.g. in NGC 4151 
\citealt{2011ApJ...729...75W,2011ApJ...736...62W,2011ApJ...742...23W}; Mrk 
573, \citealt{2012ApJ...756...39P}); in the nearby CT AGN NGC 3393, it has led 
to the discovery of two CT nuclei, with 150 pc separation 
\citep{2011Natur.477..431F}. Our new look at the nuclear region of Arp 220, 
has resulted in the discovery of two sources in the 6-7 keV Fe-K band, strongly 
suggestive of CT nuclei. These sources are spatially coincident with the near-IR 
and radio positions. Below we discuss our technique and results.

\section{Data Analysis}\label{data}
Arp 220 was observed by \textit{Chandra} on 2000 June 24 for 57 ks (Obs. ID
869, PI: Clements). Level 2 event data were retrieved from the \textit{Chandra}
Data Archive\footnote{\href{http://cda.harvard.edu/chaser}
{http://cda.harvard.edu/chaser}} and reduced with the CIAO
(\citealt{2006SPIE.6270E..60F}) 4.4 software and the \textit{Chandra}
Calibration Data Base (\textsc{caldb}) 4.5.3, adopting standard procedures.
After excluding time intervals of background flares exceeding \(3\sigma\) with 
the \textsc{lc\_sigma\_clip} task, we obtained a low-background exposure 
time of \(\sim 56\) ks. The nucleus has no significant pile up, as measured by 
the CIAO \textsc{pileup\_map} tool.
Imaging analysis was performed without pixel randomization to take advantage 
of the telescope dithering in event positioning and with the sub-pixel event 
repositioning (SER) procedure \citep{2003ApJ...590..586L}. We used a pixel 
size 1/4 of \(0.492"\), the native \textit{Chandra}/ACIS detector pixel
\citep[see, e.g.,][]
{2004ApJ...615..161H,2007ApJ...657..145S,2010ApJ...708..171P,2011ApJ...729...75W}.

{Using the same Orion ACIS-S data as in the calibration of 
\citet{2003ApJ...590..586L}, we find a significant \(\Delta=50\%\) 
(improvement in PSF FWHM as defined in \citealt{2003ApJ...590..586L}) 
from sub-pixel repositioning for an on-axis source at 6-7 keV (\(\Delta=70\% 
\) at \(\sim 2\) keV because of the narrower PSF). Most of the imaging improvement is from sub-pixel binning, which uses the 
sampling of the PSF by the well characterized spacecraft dither motion. Because
of the similarly `peaked' inner PSF this is similarly effective at 2 and 6 keV.}

\begin{figure*}
\centering
\includegraphics[scale=0.9]{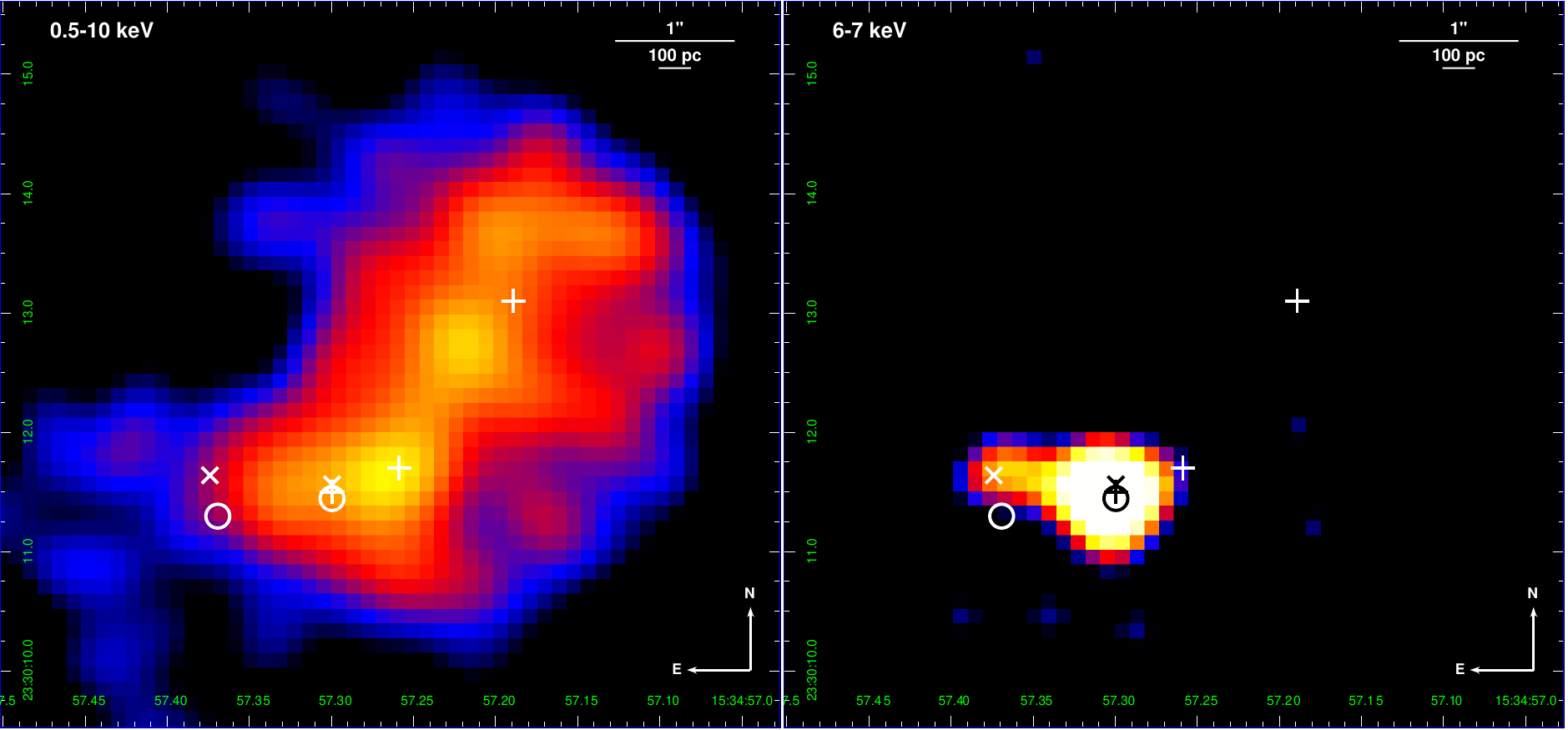}
\caption{(Left frame) Full-band (\(0.5-10\) keV) ACIS image with sub-pixel 
binning (1/4 of the native pixel size) and 2X2 FWHM gaussian filter smoothing; 
pluses (+) represent the \textit{Chandra} hard X-ray sources reported by 
\citet{2002ApJ...581..974C}, crosses (\(\times\)) represent NICMOS near-
infrared sources reported by \citet{1998ApJ...492L.107S}, and circles 
(\(\Circle\)) represent radio \textit{VLA} 6 cm peaks reported by 
\citet{1995ApJ...454..745B}, respectively. The colors of the symbols are chosen 
for visibility. (Right frame) Same of left frame but in the narrow \(6-7\) keV 
band.}\label{fullband}
\end{figure*}

The resulting full band (0.5-10 keV) ACIS image is presented in the left panel 
of Figure \ref{fullband}. This figure shows complexity in the central region of 
Arp 220, but there is no X-ray feature that can be  univocally associated with 
the radio/IR nuclei. Instead, narrow band imaging (6-7 keV containing the Fe-K 
lines) reveals these hidden nuclei (Figure \ref{fullband}, right panel). The only 
sources of emission in this spectral band are localized in two regions separated 
by \(\sim 1"\) (corresponding to \(\sim 365\) pc at a distance of 76 Mpc) and 
co-located with the NIR {\(2.2\,\mu\mbox{m}\)} 
\citep{1998ApJ...492L.107S} and radio nuclei \citep{1995ApJ...454..745B}. 
Following \citealt{1998ApJ...492L.107S}, we have shifted the \textit{VLA} 6 cm 
sources in the NE direction by \(\sim 0.13"\) in order to match the position of 
{the western lobe with the western nucleus. We note that the eastern 
radio lobe results somewhat dislocated from eastern narrow-band nucleus. 
However, due to the low counts in this region, the location of this emission is 
consistent within uncertainties with the radio lobe. Under this assumptions,} 
radio, NIR and Fe-K nuclei are consistent within astrometric uncertainties. 
{Deeper \textit{Chandra} X-ray observations are needed for convincingly 
evaluate if the position of the E nucleus is in better agreement with the IR or 
radio position.} We note that the extension of the narrow-band emission in the 
east direction is not due to PSF asymmetries, as indicated by the 
\textsc{make\_psf\_asymmetry\_region} tool that shows PSF artifacts in the 
north-west direction.

\begin{figure}
\centering
\includegraphics[scale=0.45]{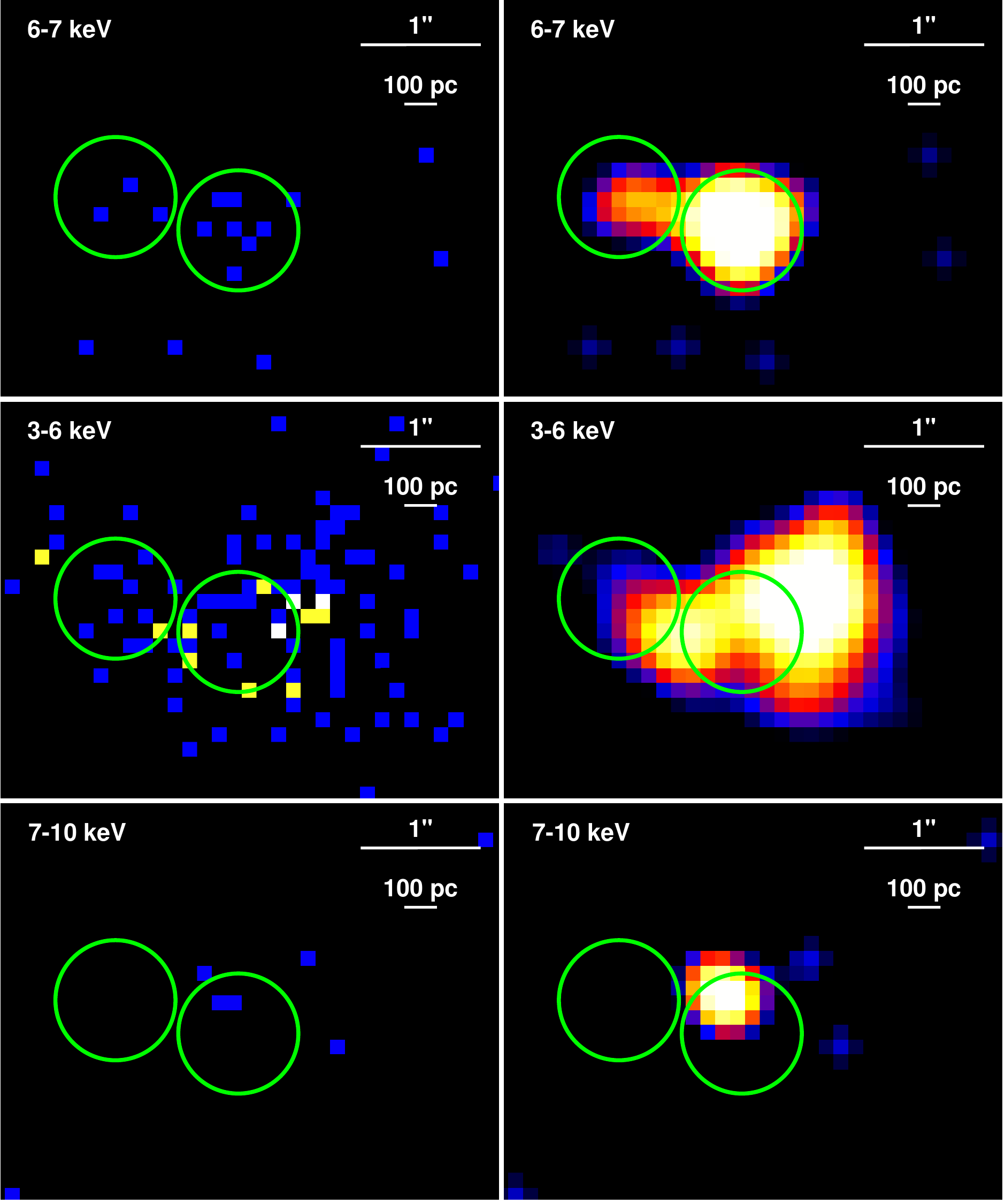}
\caption{(Upper panel) Left frame: narrow band (6-7 keV) ACIS image with 
sub-pixel binning (1/4 of the native pixel size). In the color scale is blue 
corresponds to 1 count, yellow to 2 counts, and white to \(>2\) counts. The 
large circles are our count extraction areas.
Right frame: same as left frame but with a 2X2 pixel FWHM gaussian filter 
smoothing applied. (Middle panel) Same as upper panel but in the 3-6 keV 
band. (Lower panel) Same as upper panel but in the 7-10 keV 
band.}\label{hardbands}
\end{figure}

We extracted counts from the 0''.5 circles centered at RA=15:34:57.252 
DEC=+23:30:11.64 (W) and RA=15:34:57.326 DEC=+23:30:11.84 (E) (Figure 
\ref{hardbands}, upper panel); \textit{Chandra} absolute astrometric 
uncertainty is 0".6. We find in the 6-7 keV band 12 counts associated with the 
Western nucleus (W), and 3 counts associated with the Eastern nucleus (E). Note 
that the background emission in this band is 0.01 counts in the same area of 
the extraction regions, so even a 3 counts detection is a highly significant 
source (\(P<10^{-6}\) of chance detection, corresponding to a 
\(\sim 5\sigma\) Gaussian significance).

{Given the energy dependent grade branching ratio for BI ACIS CCD 
\citep{2003ApJ...590..586L}, and since sub-pixel repositioning is more uncertain for 1 pixel (GRADE=0) events than for 2 pixel (GRADE=2,3,4) and 4 pixel (GRADE=6) events, we checked the event grade distribution in W and E regions to ensure reliability of events position with SER. The majority of W region events are 2 and 4 pixel, while no 1 pixel event is found in E region. Therefore sub-pixel analysis improves the positioning of these events.}

The flux in this band, however, is highly uncertain due to the low counts. In 
this case, approximate narrow-band model-independent fluxes can be 
estimated with the \textsc{CIAO aprates} tool, to yield 
\({0.7}_{-0.4}^{+0.5}\times{10}^{-14}\mbox{ erg}\mbox{ cm}^{-2}\mbox{ 
s}^{-1}\) (E) and \({1.9}_{-0.6}^{+0.8}\times{10}^{-14}\mbox{ erg}\mbox{ 
cm}^{-2}\mbox{ s}^{-1}\) (W). These correspond to 6-7 keV observed 
luminosities \({5.2}_{-2.5}^{+3.7}\times{10}^{39}\mbox{ erg}\mbox{ s}^{-1}\) 
(E) and \({13.8}_{-4.7}^{+5.9}\times{10}^{39}\mbox{ erg}\mbox{ s}^{-1}\) (W).

The locations of these unique emission regions strongly argue for an 
identification of these sources with the nuclei of the merging galaxies. 
Although the W source is positionally coincident with the X-4 hard X-ray
source reported by \citet{2002ApJ...581..974C} (see Figure \ref{fullband}), 
the sub-pixel narrow band imaging suggests a different picture. Figure
\ref{hardbands} (middle panel) shows the 3-6 keV band image, where the most 
prominent source is found to the west of both W and of the reported X-4 
position \citep{2002ApJ...581..974C}. Figure \ref{hardbands} (lower panel) 
shows the emission in the 7-10 keV band, suggesting hard continuum 
emission from the W nucleus (3 cts) with \(P\sim 10^{-5}\) of chance 
detection. The \textsc{CIAO aprates} tool gives in this region \({7}_{-3}^{+5}\) 
cts, while we derive a firm 2 counts upper limit in the E nucleus region. These 
images suggest the presence of highly obscured CT AGN in the nuclei.

The narrow-band images (Figure \ref{hardbands}) also show that the 
continuum emission from the nuclei is contaminated by unrelated sources even 
at \textit{Chandra} resolution. Nevertheless, we attempted a spectral 
characterization of the emission, extracting 3-8 keV spectra from the two 
circular regions (0.5" radius) indicated in Figure \ref{hardbands} with the 
\textsc{CIAO specextract} task, applying point-source aperture correction. We 
then fitted simultaneously the two spectra employing the Cash statistic. We 
used a model typical of CT AGN emission \citep{2006ApJ...648..111L}, 
comprising an absorption component fixed to the Galactic value 
\({3.9}\times{10}^{20}\mbox{ cm}^{-2}\), a pure neutral reflection component 
(\textsc{pexrav}) with a spectral index fixed to 1.8 and a gaussian Fe-K line. We 
used both \textsc{XSPEC} (ver. 12.7.1) and 
\textsc{Sherpa}\footnote{\href{http://cxc.harvard.edu/sherpa}
{http://cxc.harvard.edu/sherpa}} with identical results. The extracted spectra 
and the best-fit parameters are presented in Figure \ref{spectra} and Table 
\ref{table2}\footnote{In the following, errors correspond to the 
\(1\)-\(\sigma\) confidence level for one parameter of interest.}, respectively.

Given the contamination of the continuum emission by non-nuclear sources, 
our estimate of the nuclear continuum luminosity is an upper limit, and the Fe-
K EWs must be considered as lower limits. The spectral analysis detects Fe-K 
line features in both regions, with comparable EW (1.1 keV and 0.9 keV in W 
and E region, respectively). As expected from the imaging, the Fe-K line is 
more luminous in the W nucleus (\(5.7\times{10}^{39}\mbox{ erg}\mbox{ 
s}^{-1}\)) with respect to the E  (\(1.9\times{10}^{39}\mbox{ erg}\mbox{ 
s}^{-1}\)); we note that these values are consistent within errors with the 
narrow-band (6-7 keV) luminosities obtained with \textsc{aprates} tool. As 
already discussed, the \textit{observed} 2-10 keV luminosities, 
\(3.2\times{10}^{40}\mbox{ erg}\mbox{ s}^{-1}\) (W) and
\(1.3\times{10}^{40}\mbox{ erg}\mbox{ s}^{-1}\) (E), should be considered as 
upper limits because of contamination.

Our detection of large Fe-K EWs from both nuclei may seem at odds with the 
results of \citet{2011ApJ...729...52L}, who did not detect Fe-K emission using 
the same Chandra ACIS data. To investigate this discrepancy, we repeated the 
analysis of the entire central emission of Arp 220, following the procedure of 
these authors. Using a circular 4.5" radius count extraction region centered at 
RA=15:34:57.194 DEC=+23:30:12.40, and their fitting model, we confirm 
their results (Table \ref{table}, left column of results). However, the authors 
fixed the Fe-K line rest-frame energy fixed at 6.4 keV. If, instead, we allow the 
gaussian line rest-frame energy to vary, we detect a line at \({6.61}\pm{0.04}\) 
keV with and equivalent width of \({0.73}_{-0.30}^{+0.60}\) keV, compatible 
with the \textit{XMM-Newton} detection at \(6.72\pm 0.5\) keV 
\citep{2005MNRAS.357..565I}; all the other model parameters remain 
unchanged (see the right column in Table \ref{table}).

As demonstrated by Figures \ref{fullband} and \ref{hardbands} there 
is no 6-7 keV emissions in regions outside the nuclei, even when the overall 
X-ray emission is more prominent. We have also extracted  the spectrum of the 
softer luminous regions west of the W nucleus shown in Figure \ref{hardbands} 
(middle-right panel) centered at RA=15:34:57.207 DEC=+23:30:11.84, and 
find no evidence of line emission.

{We analyzed the four archival \textit{XMM-Newton} observations of Arp 
220 (the two discussed in \citealt{2005MNRAS.357..565I}, and two new ones 
performed in 2005) in order to test the possible contribution of a neutral iron 
emission line to the 6-7 keV emission we see in Figure \ref{hardbands}. The 
data were reduced following a standard procedure, analogous to the one 
described by \citet{2005MNRAS.357..565I}. The results are also in agreement: 
in a continuum plus single line model we obtain a best fit peak energy 
\(E=6.65\pm 0.04\mbox{ keV}\). However, if we fit the data with two lines with 
fixed peak energies \(E_1=6.4\mbox{ keV}\) and \(E_2=6.7\mbox{ keV}\), we 
obtain the results shown in Figure \ref{contours}: a neutral component 
accounting for up to 40\% of the observed line flux cannot be ruled out at a 
90\% confidence level.}

{If we then consider that 40\% of the Fe-K line flux we estimate from 
\textit{Chandra} spectra is due to Fe-K\(\alpha\) neutral 6.4 keV emission line, 
the 2-10 keV \textit{emitted} luminosities inferred from the Fe-K luminosities 
are \(1.0\times{10}^{42}\mbox{ erg}\mbox{ s}^{-1}\) (W) and 
\(0.4\times{10}^{42}\mbox{ erg}\mbox{ s}^{-1}\) (E). We note that these 
corrections are calibrated on ``standard" obscured Seyfert galaxies, with an 
X-ray reflection efficiency of a few percent \citep{2006ApJ...648..111L}. Hard 
X-ray observation of ULIRGs have demonstrated that on average this efficiency 
is much lower for these sources 
\citep{2011MNRAS.415..619N,2012AAS...22040903T}. Consequently, the 
intrinsic X-ray luminosity of the two AGN detected here could be significantly 
higher. Considering the values for a standard reflection efficiency, the inferred 
X-ray luminosity is at least a factor of 3 higher than that expected from a pure 
starburst with the bolometric luminosity of Arp 220 
\citep{2003A&A...399...39R}.}

\section{Discussion}\label{discussion}
 
The Chandra ACIS sub-pixel narrow-band imaging of the central region of the 
ULIRG merger Arp 220 provides compelling evidence of two CT AGNs, in both 
nuclei of the merging galaxies. Within the central 5''x5'', there are \textit{only} 
two sources detected in the 6-7 keV band, containing the Fe-K lines. Although 
the E nucleus is detected with only 3 counts, the very low field background 
(\(\sim 0.01\)) makes this a \(5 \sigma\) detection (Section \ref{data}). The 
centroids of these sources, \(\sim 1\)'' apart,  are consistent, within 
\textit{Chandra} astrometric uncertainty of 0.6'', with the position of the two 
NIR nuclear clusters identified by \citet{1998ApJ...492L.107S}, and each 
coincident with a 6 cm VLA radio source \citep{1995ApJ...454..745B} (Figure 
\ref{fullband}). We note that the W source is also consistent with the variable 
radio sources reported by \citet{2012A&A...542L..24B}. We stress that 
\textit{no emission} in the 6-7 keV band is detected from other parts of the 
central region of Arp 220, even where the diffuse emission from the starburst 
is most intense. While the detections are highly significant for both nuclei, the 
fluxes are more uncertain, given the small number of detected photons (see 
Section \ref{data}).

The spectral analysis of both the central 4.5'' region, and of the individual E 
and W nuclei, results in the detection of emission lines. The rest energy of the 
line is larger than the 6.4 keV of the K\(\alpha\) line, 
{and suggests a contribution from 6.7 keV shock-ionized Fe 
\textsc{xxv} line as concluded by \citet{2005MNRAS.357..565I}, which could be 
associated with the starburst-induced shock (see e.g. the strong extended Fe 
\textsc{xxv} emission in NGC 6240 \citealt{wang2013}). However, if we 
\textit{assume} that both 6.4 keV and 6.7 keV lines are present in the 
spectrum, we obtain an acceptable fit to the \textit{XMM-Newton} data which 
allow for 40\% 6.4 keV contribution. The statistics, however, do not allow us to 
disentangle these two contributions to the observed emission in 
\textit{Chandra} data.} However, as discussed in Section \ref{data}, the line 
emission is only connected with the nuclear region, not with the more 
extended starburst. Although the uncertainties are large, we find large \(\sim 
1\) keV equivalent widths for the Fe-K lines; {simulated data show that 
shock-ionized gas can yield comparable EWs for these lines, but would also 
yield continuum contribution much higher than we observe.}

\begin{figure}
\centering
\includegraphics[scale=0.34]{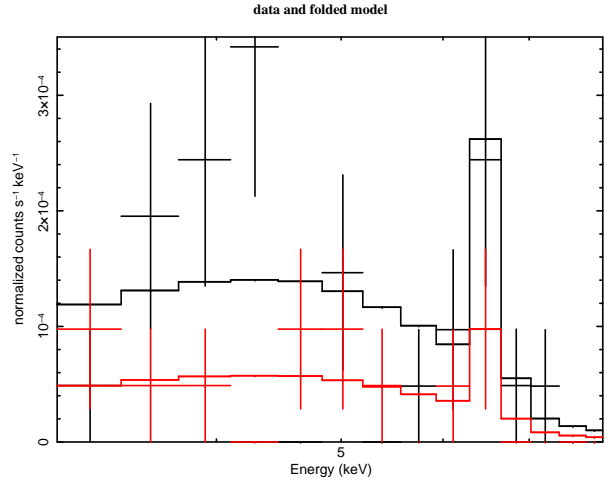}
\caption{Best fit to spectra extracted from regions shown in Figure 
\ref{hardbands}. The spectrum from W region is shown in black and the E 
region in red.}\label{spectra}
\end{figure}

\begin{figure}
\centering
\includegraphics[scale=0.34]{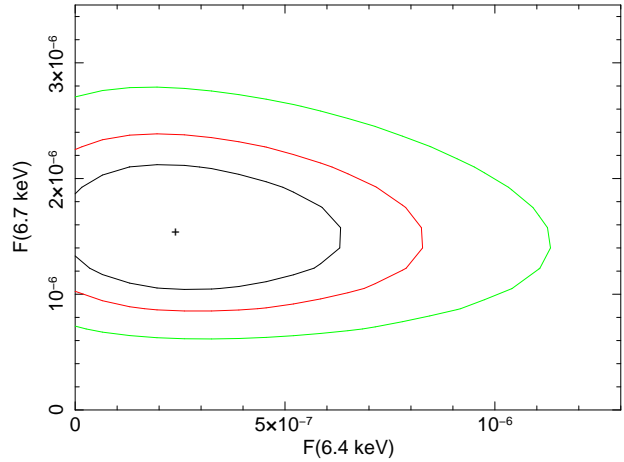}
\caption{Contour plot for the relative flux of the 6.4 and 6.7 keV lines in a fit where both energies are fixed. The contours represent a probability of 68\%, 90\% and 99\%. Even if a solution with a single 6.7 keV line is possible, a contribution of up to 40\% by a neutral line is compatible with the data at a 90\% confidence level.}\label{contours}
\end{figure}

Besides a definite detection of hard photons (\(>7\) keV) from the W nucleus, 
which is also consistent with it being a CT AGN, the continuum emission from 
the nuclear regions is not easy to establish because of contamination from the 
surrounding emission in the 3-6 keV band. In particular, the W nucleus is 
significantly contaminated by the softer emission of a source or extended 
emission area nearby, which could be connected with the starburst 
phenomenon. This complex circumnuclear emission impedes the measurement 
of the nuclear continuum by means of spectral fitting. Any determination of the 
nuclear continuum with lesser spatial resolution data will give an even higher 
overestimation of its strength. Considering the above caveats, we can only set 
an upper limit of \(< 2\%\) to the ratio between the observed and the inferred 
intrinsic luminosities for the two nuclear sources (see Section \ref{data}). This 
limit is consistent with the ratio between the observed luminosities of less 
obscured and CT AGNs \citep{1998A&A...338..781M}. The spectra of the two 
nuclear sources presented in Figure \ref{spectra} are in fact typical of CT AGNs, 
with observed Fe-K EW \(\gtrsim 1\) keV.

\begin{figure}
\centering
\includegraphics[scale=0.34]{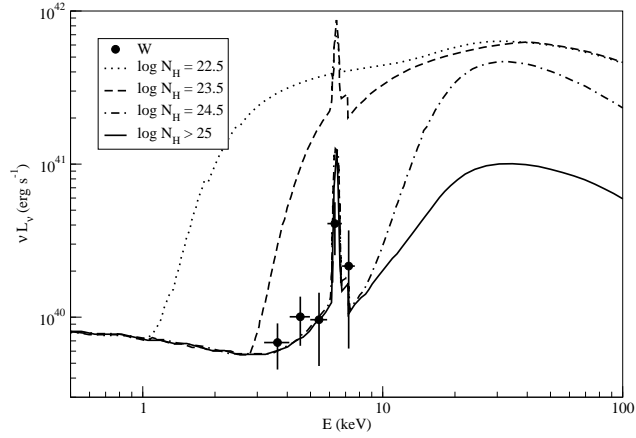}
\caption{Rest frame SEDs of the W nuclear source in Arp 220 (black circles) 
compared with simulated AGN X-ray spectra with different absorbing column 
densities \citep{2007A&A...463...79G}.}\label{sed}
\end{figure}

The possibility of Arp 220 harboring one heavily obscured AGN has been 
discussed by \citet{2001MNRAS.326..894I,2005MNRAS.357..565I} with 
\textit{Beppo}SAX and \textit{XMM-Newton} data; these authors conclude that, 
due to lack of hard emission above 10 keV, the CT AGN should be enshrouded 
in absorbing clouds with column density exceeding \({10}^{25}\mbox{ 
cm}^{-2}\) and covering factor close to unity. Even ignoring the contamination 
of the continuum by extra-nuclear sources, the low counts in the two regions 
do not allow us to estimate absorption through spectral fitting. We can 
however compare the X-ray spectral energy distribution (SED) of the W source, 
for which we detect the hard nuclear continuum, with simulated spectra of 
obscured AGNs from \citet{2007A&A...463...79G} {(see Figure 
\ref{sed})}. For this purpose we use only the spectrum at energies \(> 5\) keV, 
where the continuum is low and less contaminated by the nearby source 
{(see Figure \ref{spectra})}. The SED of the W nucleus is consistent with 
emission absorbed by \(N_H \sim{10}^{24.5}\mbox{ cm}^{-2}\). For 
comparison we extracted the spectrum from the bright region visible in Figure 
\ref{hardbands} (middle-right panel) west of W region; besides not showing 
any appreciable Fe-K feature, it is also compatible with absorbing column 
densities \(<{10}^{22}\mbox{ cm}^{-2}\).

The two nuclei have radio fluxes of 112 (W) and 88.2 (E) mJy 
\citep{1995ApJ...454..745B}, and X-ray to optical ratios \(\alpha({ox}) 
\approx 2.75\). From the inferred 2-10 keV emitted luminosities of the two 
nuclear sources we evaluate bolometric luminosities assuming typical AGN 
SEDs \citep{1994ApJS...95....1E,2002ApJ...565L..75E}, and the X-ray reflection 
efficiency of Seyfert galaxies, as discussed in Section \ref{data}. The estimated 
AGN bolometric luminosities, which should be regarded as lower limits, are 
{\(\sim 5.2\times{10}^{43}\mbox{ erg}\mbox{ s}^{-1}\) (W) and \(\sim 
2\times{10}^{42}\mbox{ erg}\mbox{ s}^{-1}\) (E)}. These represent only a few 
percent of Arp 220 bolometric luminosity, confirming that overall the emission 
of Arp 220 is dominated by the starburst component. This result is in broad 
agreement with the estimates from the mid-IR spectroscopy 
\citep{2009ApJS..182..628V,2010MNRAS.405.2505N}. 
Lower limits on associated BH masses can be evaluated assuming Eddington 
limited accretion (with a standard 10\% accretion rate to luminosity conversion 
efficiency), yielding \(M \sim 3\times{10}^5 M_{\astrosun}\) (W) and \(\sim 
1\times{10}^5 M_{\astrosun}\) (E).
 
Our results add to the evidence of CT AGNs arising as a result of the later 
stages of the merging evolution in galaxies, which may trigger accretion onto 
the supermassive black holes  \citep[e.g.,][]{2012ApJ...748L...7V}. After NGC 
6240 \citep{2003ApJ...582L..15K} and NGC 3393 \citep{2011Natur.477..431F}, 
Arp 220 provides the third clear case of the occurrence of this phenomenon in 
the near universe. As NGC 6240 \citep{2003ApJ...582L..15K}, Arp 220 is a 
highly disturbed system of galaxies engaged in a major merging interaction.
The physical projected separation of the CT nuclei is \(\sim 670\) pc in NGC
6240 and \(\sim 370\) pc in Arp 220, suggesting that the latter may be in a 
more advanced stage of merging. The third case, the apparently regular early-
type spiral NGC 3393, with two CT nuclei separated by 150 pc 
\citep{2011Natur.477..431F}, suggested instead a more evolved merger, or 
perhaps a minor merger of unequal size galaxies. Interestingly, following the 
discovery of the double nuclear X-ray source, evidence of a merger past has \
surfaced in the optical spectra of this galaxy \citep{2012MNRAS.tmp..247C}.

\section{Conclusions}\label{conclusions}
We have found compelling evidence for two CT AGNs, associated with the 
nuclei of the merging galaxies in the nearest ULIRG Arp 220, making this 
galaxy the third case of detection of close pair CT AGNs (a few 100 pc apart), 
after NGC 6240 and NGC 3393. These nuclei are the sole regions of significant 
6-7 keV emission in the central 2 kpc of Arp 220. The data are consistent with 
Fe-K line emission although at an energy possibly higher than 6.4 keV of the 
Fe-K\(\alpha\) line, {suggesting substantial contribution from Fe 
\textsc{xxv}} emission lines \citep[see also][]{2005MNRAS.357..565I}. 
{Our analysis of the entire \textit{XMM-Newton} dataset confirms the 
presence of Fe \textsc{xxv} emission line, but allows 40\% of the narrow-band 
emitted flux form the neutral 6.4 keV line.} Albeit uncertain - the spectral 
analysis of these regions suggests large \(\sim 1\) keV EWs. The W nucleus 
was also detected at hard (\(>7\) keV) energies implying absorption \(N_H 
\sim{10}^{24.5}\mbox{ cm}^{-2}\); at such energies no continuum emission 
was detected from the E nucleus. Our results are consistent with previous 
multi-wavelength indications of nuclear activity in Arp 220 (see Section 
\ref{intro}), and strengthen the evolutionary association of merging and nuclear 
activity in galaxies \citep[e.g.,][]{2008ApJ...682L..13H,2012ApJ...748L...7V}. 
Based on the Fe-K detections, we infer lower limits on the bolometric 
luminosity of the AGNs of \(5.2\times{10}^{43}\mbox{ erg}\mbox{ s}^{-1}\) for 
the W AGN, and \(\sim 2\times{10}^{42}\mbox{ erg}\mbox{ s}^{-1}\) for the E 
AGN. These are a few percent of the total ULIRG bolometric luminosity, 
confirming that overall the emission of this source is dominated by the 
starburst component, as estimated from the mid-IR spectroscopy 
\citep{2009ApJS..182..628V,2010MNRAS.405.2505N}. 

These results have only been possible because of the unmatched 
\textit{Chandra} spatial resolution, and the use of sub-pixel imaging in narrow 
spectral bands, which expose the telltale Fe-K emission of CT AGNs, and give 
us a clear picture of the nuclear surroundings. Although the Fe-K band 
detections of the two nuclei are highly significant, the paucity of photons 
results in large uncertainties for all derived quantities. A significantly longer 
\textit{Chandra}/ACIS exposure will be needed to firmly measure  the emission 
parameters of the two nuclei.

\acknowledgments
{We acknowledge useful comments and suggestions by our anonymous referee.}
{We thank Jonathan McDowell for data analysis suggestions.}
{This work is supported by NASA grant GO1-12125A.}
{We acknowledge support from the CXC, which is operated by the Smithsonian 
Astrophysical Observatory (SAO) for and on behalf of NASA under Contract 
NAS8-03060.}
{This research has made use of data obtained from the Chandra Data Archive, 
and software provided by the CXC in the application packages CIAO and 
Sherpa.}
{This research has made use of Iris software provided by the US Virtual 
Astronomical Observatory, which is sponsored by the National Science 
Foundation and the National Aeronautics and Space Administration.}

\newpage

\begin{table}
\centering
\caption{Spectral fitting for the iron regions}\label{table2}
\begin{tabular}{lcc}
\hline
\hline
& W region & E region \\
Band & \multicolumn{2}{c}{Net counts (error)} \\
 3-8 keV & 30(5) & 7(3) \\
 6-7 keV & 12(3) & 3(2) \\
\hline
Model parameter & \multicolumn{2}{c}{Best-fit value} \\
\(E_{Fe-K}\) (keV) & \({6.65}_{-0.05}^{+0.06}\) & \({6.56}_{-0.10}^{+0.11}\) \\
\(F_{Fe-K} ({10}^{-6}\mbox{ cm}^{-2}\mbox{ s}^{-1})\) & \({0.75}_{-0.37}^{+0.50}\) & \({0.25}_{-0.19}^{+0.31}\) \\
\(L_{Fe-K} ({10}^{39}\mbox{ erg}\mbox{ s}^{-1})\) & \({5.70}_{-2.79}^{+3.77}\) & \({1.87}_{-1.46}^{+2.32}\) \\
EW (keV) & \({1.10}_{-0.75}^{+0.62}\) & \({0.90}_{-0.79}^{+0.85}\) \\
C-stat (d.o.f.) & \multicolumn{2}{c}{211.10(676)} \\
\hline
\(L_{2-10\mbox{ keV}} ({10}^{40}\mbox{ erg}\mbox{ s}^{-1})\) & \({3.17}_{-0.67}^{+0.65}\) & \({1.26}_{-0.36}^{+0.30}\) \\
\hline
\hline
\end{tabular}
\end{table}

\begin{table}
\centering
\caption{Spectral fitting for the central 4.5" region}\label{table}
\begin{tabular}{lcc}
\hline
\hline
Band & \multicolumn{2}{c}{Net counts (error)} \\
 0.5-10 keV & \multicolumn{2}{c}{565(24)} \\
 6-7 keV & \multicolumn{2}{c}{17(4)} \\
\hline
Model parameter & \multicolumn{2}{c}{Best-fit value} \\
\(N_H (\mbox{ cm}^{-2})\) & \({0.76}_{-0.22}^{+0.14}\) & \({0.73}_{-0.22}^{+0.15}\) \\
kT (keV) & \({0.91}_{-0.06}^{+0.07}\) & \({0.91}_{-0.06}^{+0.07}\) \\
\(\Gamma\) & \({0.95}_{-0.28}^{+0.24}\) & \({1.05}_{-0.27}^{+0.22}\) \\
\(E_{Fe-K}\) (keV) & 6.4* & \({6.61}\pm{0.04}\) \\
\(F_{Fe-K} ({10}^{-6}\mbox{ cm}^{-2}\mbox{ s}^{-1})\) & \(<0.32\) & \({0.87}_{-0.36}^{+0.44}\) \\
\(L_{Fe-K} ({10}^{39}\mbox{ erg}\mbox{ s}^{-1})\) & \(<3.64\) & \({6.61}_{-2.73}^{3.34}\) \\
EW (keV) & \(<0.42\) & \({0.73}_{-0.30}^{+0.60}\) \\
C-stat (d.o.f.) & 128.94(130) & 123.42(129) \\
\hline
\(L_{2-10\mbox{ keV}} ({10}^{40}\mbox{ erg}\mbox{ s}^{-1})\) & \({7.80}_{-1.48}^{+0.66}\) & \({7.89}_{-1.41}^{+0.43}\) \\
\hline
\hline
\end{tabular}
\end{table}

\end{document}